# Coherent transfer of optical orbital angular momentum in multi-order Raman sideband generation

J. Strohaber[1,*], M. Zhi[1], A. Sokolov[1], A. A. Kolomenskii[1], G. G. Paulus[1,2] and H. A. Schuessler[1]

[1]*Texas A&M University, Department of Physics and Institute for Quantum Studies, College Station, TX 77843-4242, USA*
[2]*Institut für Optik und Quantenelektronik, Max-Wien-Platz 1, 07743 Jena, Germany*

*Corresponding author: jstroha1@gmail.com*

Experimental results from the generation of Raman sidebands using optical vortices are presented. By generating two sets of sidebands originating from different locations in a Raman active crystal, one set containing optical orbital angular momentum and the other serving as a reference, a Young's double slit experiment was simultaneously realized for each sideband. The interference between the two sets of sidebands was used to determine the helicity and topological charge in each order. Topological charges in all orders were found to be discrete and follow selection rules predicted by a cascaded Raman process.

Laguerre-Gaussian $LG_p^\ell$ beams have generated considerable interest over a broad range of disciplines due to applications such as optical tweezers and spanners, super resolution imaging, and entangled quantum states [1]. These unique beams of light, known as Optical vortices (OVs), are solutions to the paraxial wave equation (PWE) and carry a physical quantity known as optical angular momentum. The concept of angular momentum in optics is not unique to OVs; however, since these beams are *eigensolutions* of the PWE they carry a sharply defined amount of optical orbital angular momentum (OAM) per photon. We are interested in investigating the interaction of OAM containing beams with matter to elucidate the role that this quantity plays in ultrafast intense-field processes, which still remains experimentally largely unexplored territory.

The generation of coherent radiation spanning a broad spectral range is currently an active field of research and is used in the production of ultrashort pulses of radiation. In addition to broader bandwidths, creating ultrashort pulses requires shorter wavelengths. Two methods have emerged in this endeavor: cascaded Raman sideband generation and High Harmonic Generation (HHG) [2]. In the UV/XUV spectral region, radiation from HHG have demonstrated the production of single isolated ~150 as pulses using a now well-developed technique based on optical gating [3]. Compared to Raman sideband generation, HHG is inherently inefficient having a conversion efficiency on the order $10^{-6}$. In the vis-UV region, sideband generation has taken center stage. Sideband generation is highly efficient compared to HHG, converting one out of ten photons into higher energy photons [2]. An additional advantage of producing radiation in this spectral region is the availability of a plethora of optical devices which can be utilized to modify the both spectral and spatial content of the resulting radiation. For example, H. S. Chan *et. al.,* have synthesized various temporal waveforms by manipulating the phases and amplitudes of the individual sidebands [4].

While sideband generation from the parametric coupling between a pump (P) and Stokes (S) beam presents an attractive prospect for attosecond physics, currently they are of limited use since the shortest pulses that have been experimentally realized are from a train of 1.4 fs pulses, and single isolated 2.7 fs pulses using impulsive techniques [5]. In this letter, results from the generation of Raman sidebands using OAM containing beams are presented. Manipulation of the transverse spatial profiles of the pump and Stokes beams offers an extra degree of freedom in which ultrashort OVs can be produced or converted to different wavelengths.

In nonlinear optical phenomena that are second order in the field, nonlinear polarizations give rise to radiation containing contributions from the second harmonic, the sum and difference frequencies, and optical rectification [6]. Using various radiation sources, nonlinear media and experimental configurations; sum-frequency and second harmonic generation [Fig. 1(a)] using OVs have been investigated by a number of authors [7—9]. In these experiments the net topological charge of the resulting beams have been found to follow basic arithmetical operations $\ell_3 = \ell_1 + \ell_2$, and these processes have been viewed by some authors as a prototypical optical processor [8]. Difference-frequency generation [Fig. 1(b)] with OAM containing beams has also been investigated under spontaneous and stimulated processes. On the microscope scale spontaneous down conversion experiments have demonstrated the production of entangled photons pairs which conserve OAM [9]. Conversion of topological charge has also been observed in stimulated down conversion experiments [10].

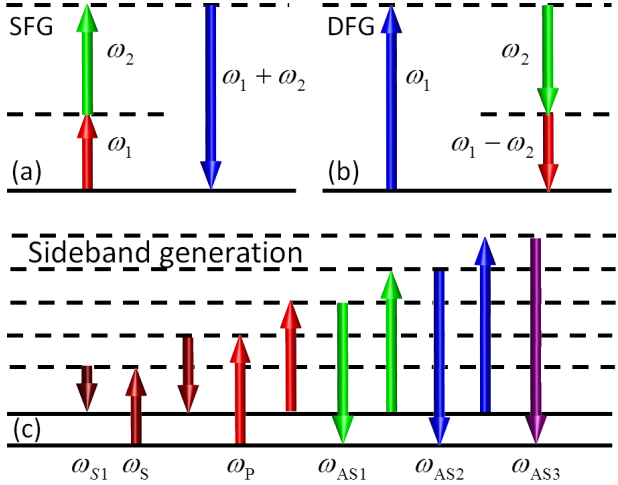

Fig. 1. Energy level diagrams for (a) sum-frequency generation, (b) difference-frequency generation, and (c) cascaded Raman sideband generation.

In the current work, experiments were performed using a 1 kHz Ti:Sapphire (Spectra-Physics Spitfire) laser system emitting radiation with a center wavelength of 800nm, a pulse duration of ~50 fs and an output power of 1 mJ per pulse. This radiation was sent into a four-port Michelson interferometer (not shown) so that the two output beams could be used to generate two independent sets of sidebands. Optical vortices of topological charge $|\ell|=1$ were produced in one of the output ports by positioning a spiral phase plate (SPP)[HoloOr VL-209-M-Y-A] directly after the first beam splitter. Because the SPP was constructed using 16 different step heights and due to a mismatch between the final step height and radiation other than 800 nm, all radiation is not expected to be in the desired mode $LG_0^1$ [1,11]. Immediately following the SPP, only the spatial phase of the beam is modified, leaving the transverse amplitude distribution the same as the incoming radiation. For a well-designed SPP, it is expected that most of the radiation resides in modes having $\ell=0$ and $p=0$ with the remainder in modes having different radial mode numbers. Ultimately, visual inspection of the OV-reference interferogram on a white screen following the interferometer showed a single bifurcation having sharp contrast as expected from a pure $LG_0^1$ mode.

The output of the four-port interferometer was sent into the beam crossing setup shown in Fig. 2. Input radiation was split by a beam splitter (BS), which directs radiation into each arm of the setup. The variable arm consisted of a translational stage (T) and three mirrors M1, M2, M3; the fixed arm had a total of four mirrors (M4—M7). By adjusting M4 and M6, radiation could be reflected from M5 (dotted line in Fig. 2) to introduced an extra reflection. In the absence of M5, the beam crossing setup is symmetric with respect to linearly polarized radiation having $\ell=0$ but not with respect to an LG beam with $\ell\neq 0$ or a circularly polarized beam: an asymmetry is caused by BS which introduces an extra reflection in the fixed arm. A 23 cm focal length lens focused radiation into a 1 mm thick lead tungstate crystal PbWO$_4$ at a crossing angle of ~ 3°. The *strong* narrow Raman line at 901 cm$^{-1}$ lies outside of the bandwidth of our radiation; however, the relatively *weaker* Raman mode at 320 cm$^{-1}$ was accessible. To generate Raman sidebands, time-delayed linear chirped pulses having sufficient chirp (experimentally found) together with a loose focusing geometry was employed [details can be found in Ref(12)]. For a fixed chirp parameter $b$, a delay between the pump and Stokes can be found such that the periodicity of the resulting pulse train can be matched to the period of the Raman mode $\Delta\omega = \omega_R = bt_d$ permitting selective excitation. This configuration has the advantage that a single radiation source can be used, and chirped pulses lessen contributions from unwanted nonlinear phenomena such as self-focusing, self-phase modulation and four-wave mixing. Additionally, the non-collinear setup eliminates the need for dispersive optics elements to separation the different sideband [12].

Charged-coupled device (Spiricon SP620U) images of the experimental results are shown in Fig. 3. Over 20 sidebands were observed, and spectral measurements of the first 4 AS orders verified selective excitation of the weaker Raman modes at 320 cm$^{-1}$. With mirror M5 in place, the reflections in both arms are balanced in the sense that an odd number of reflections occur. The first and second rows show the OVs and interferograms produced in sidebands AS2 through S2 with M5 in place. Rows 3 and 4 are those with mirror M5 removed. Relative to the pump and Stokes beams, a difference in vortex size can be observed. All OVs in row 1 have comparable sizes while those in row 3 appear to increase in size with increasing S or AS order. The similarity (difference) in size suggests identical (different) topological charges. This can be deduced because the ring size of OVs increases with topological charge $r \propto |\ell|^{1/2}$. However, this increase alone is not a sufficient indicator of the topological charge.

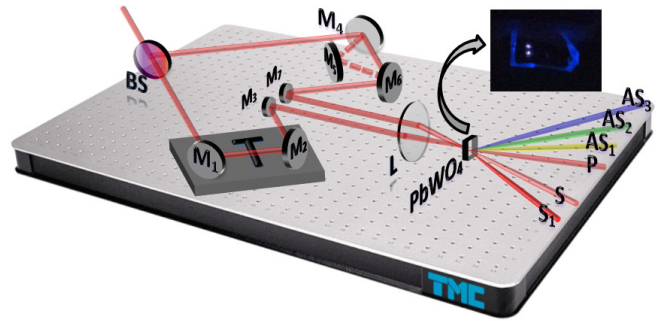

Fig. 2. Beam crossing setup. (BS) beam splitter, (M) mirrors, (T) translation stage, (L) lens, (PBWO4) lead tungstate crystal, (S) Stokes beams, (P) pump, and (AS) anti-Stokes orders.

To investigate the topological charges of the OVs in each order more definitively, a Young's double slit experiment was realized for each order simultaneously by allowing the reference beam from the four-port interferometer to propagate through the beam crossing setup along with the OV beam. The reference beam was shifted slightly in the vertical direction allow the two sets of P and S beams to cross at a different locations within the crystal (inset shown in Fig. 2) producing two sets of Raman sidebands. The inteferograms are shown in rows 2 and 4 of Fig. 3. The contrast of the inteferograms was maximized by adjusting the variable arm of the Michelson. The direction of the multifurcations give the helicity of the OVs once the sign of the angle between the interfering beams is known. For our analysis and interpretation of the data, it is sufficient to know only the relative helicity of the OVs.

The experimental results can be described by a cascaded Raman process Fig 1(c) [2]. Since the angular dependence of OVs appears in the phase along with the frequency, the topological charges in the S and AS orders can be determined in a similar fashion as one would determine the frequency of the radiation in each order. The phases of the pump and Stokes beams can be written as $\varphi_P = -\omega_P t + \ell_P \theta$ and $\varphi_S = -\omega_S t + \ell_S \theta$ respectively. From the energy diagram in Fig. 1(c), the phase of AS1 is found by subtracting the phase of the Stokes from twice the phase of the pump $\varphi_{AS1} = 2\varphi_P - \varphi_S$, and for AS2 the phase of the Stokes will be subtracted from the sum of the phases of the pump and AS1 viz., $\varphi_{AS2} = \varphi_P + \varphi_{AS1} - \varphi_S$. In the $n^{th}$ AS order, this prescription leads to a rule for the frequencies $\omega_n^{AS} = (n+1)\omega_P - n\omega_S$ and mode numbers $\ell_n^{AS} = (n+1)\ell_P - n\ell_S$. Using similar arguments, those in the S orders are found to be $\omega_n^S = (n+1)\omega_S - n\omega_P$ and $\ell_n^S = (n+1)\ell_S - n\ell_P$. Adding these two results, we find the conservation rule $\ell_n^S + \ell_n^{AS} = \ell_S + \ell_p$. Similar selection rules have been found in the case of degenerate two-wave mixing in Kerr-like nonlinear medium [13].

When both the Stokes and pump beams have the same value of $\ell$, the topological charge in all orders is $\ell_S^n = \ell_{AS}^n = \ell$, which is independent of the order. The interferograms in row 1 are consistent with this result showing bifurcations $\ell = \pm 1$ situated in the same direction. This result may be of particular interest in synthesizing custom broadband ultrashort pulses in a pure transversal mode since each *chromatic* order has the same value of $\ell$. When the Stokes and pump beams have the same value of $\ell$ but differ in sign, the bifurcations of the pump and Stokes beams face in opposite directions. The topological charge and helicities in the $n^{th}$ order consist of odd integer values $\ell_n^{AS} = -\ell_n^S = (2n+1)\ell$. Careful fringe counting analysis of the multifurcations in each order verified the predicted topological charges for the measured S and AS orders (i.e., $\pm 5, \pm 3, \pm 1, \mp 1, \mp 3, \mp 5$).

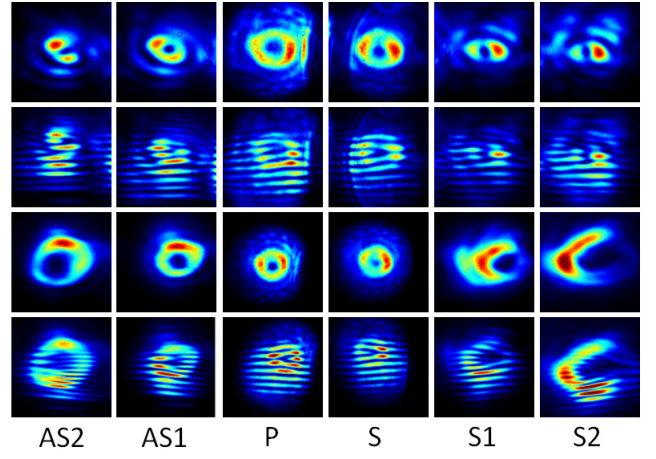

Fig. 3. Raman sideband generation with optical vortices (OV). (AS) anti-Stokes orders, (P) pump beam (S) Stokes beam and orders. Rows 1 and 2 (with M5 in place, see text) are OVs and interferograms from sideband generation with P and S having $\ell_P = \ell_S = \ell$ ($\ell = \pm 1$) and rows 3 and 4 (without M5) having $\ell_P = -\ell_S = \ell$ ($\ell = \pm 1$). The order and orientation of the multifurcation give the helicities and topological charges of the resulting OVs.

In conclusion, we have generated broadband Raman sidebands in a Raman-active crystal with the goal of synthesizing ultrashort pulses of radiation having spatially modified waveforms. In particular, we have realized the coherent transfer of OAM in the selectively excited Raman transitions in a PbWO$_4$ crystal by using a pair of time-delayed linearly chirped pulses.

This work was partially supported by the Robert A. Welch Foundation (grant No. A1546 and A1547), the National Science Foundation (NSF) (grant No. 0722800) and the U.S. Army Research Office (grant No. W911NF-07-1-0475).